\documentclass[prc,aps,groupedaddress,superscriptaddress,twocolumn,nofootinbib,showpacs,showkeys,floatfix,a4paper,10pt,amsmath,amssymb]{revtex4-1}
\usepackage{graphicx}
\usepackage{dcolumn}
\usepackage{bm}

\begin{document}

\title{Quadrupole-hexadecapole coupling in the rare earth region with 
beyond mean field correlations}

\author{R. Rodr\'{\i}guez-Guzm\'an}
\email{guzman.rodriguez@nu.edu.kz}
\affiliation{Department of Physics, School of Sciencies and Humanities, Nazarbayev 
University, 53 Kabanbay Batyr Ave., Astana 010000, Kazakhstan}

\author{L. M. Robledo}
\email{luis.robledo@uam.es}
\affiliation{%
Center for Computational Simulation, Universidad Polit\'ecnica de 
Madrid, Campus Montegancedo, 28660 Boadilla del Monte, Madrid, Spain
}%
\affiliation{Departamento  de F\'{\i}sica Te\'orica and CIAFF, 
Universidad Aut\'onoma de Madrid, 28049-Madrid, Spain}

\date{\today}

\begin{abstract} 
The roles of static hexadecapole deformation and beyond-mean-field 
quadrupole-hexadecapole configuration mixing are studied for a selected 
set of Yb, Hf, W and Os isotopes within the mass range $170 \le A \le 
202$, using the Hartree-Fock-Bogoliubov (HFB) and the two-dimensional 
Generator Coordinate Method (2D-GCM) approaches, based on the Gogny 
energy density functional. The 2D-GCM ground and excited states of the 
lighter isotopes are associated with diamond-like shapes while, for 
each isotopic chain, a region where those states correspond to 
square-like shapes  has been found below the neutron shell closure 
$N=126$. It is shown, that for the studied nuclei the quadrupole and 
hexadecapole degrees of freedom are interwoven in the ground and 
excited states up to the mass number  $A=184-188$. This structural 
evolution, encoded in the 2D-GCM collective wave functions, is 
accompanied by an enhanced prolate-oblate shape coexistence around the 
neutron number $N=116$. In agreement with previous studies, it is also 
shown that for the considered Yb, Hf, W and Os isotopes the inclusion 
of hexadecapole deformation in the ground state dynamics leads to a non 
trivial additional correlation energy comparable to the quadrupole 
correlation energy itself.

\end{abstract}


\maketitle{}

%
%
%

\section{Introduction.}
The concept of deformation in the intrinsic system plays a prominent role 
in our understanding of low-energy nuclear structure. Both 
ground and excited states of atomic nuclei exhibit a rich variety
of shapes, characterized by the corresponding multiple moments 
$Q_{\lambda \mu}$ and their 
associated deformation parameters $\beta_{\lambda \mu}$. Fingerprints
of intrinsic deformations in those ground and excited states are well
known all over the nuclear chart \cite{rs}. 
 
The axial  
$Q_{20}$ and triaxial $Q_{22}$ components of the quadrupole moment, associated with the 
deformation parameters $\beta_{20}$ and
 $\gamma$, respectively, have already received close 
scrutiny using a variety of theoretical models  
\cite{triaxial-example-1,triaxial-example-2,triaxial-example-3,Others-example-1,
Others-example-2,Others-example-3,Others-example-4,Others-example-5, triaxal-example-8,triaxal-example-9,triaxal-example-10,triaxal-example-11}. The 
impact of the $\gamma$ degree of freedom on the heights of  static inner fission
barriers in heavy and superheavy nuclei has also been considered in previous 
studies (see, for example, \cite{triaxial-example-4,triaxial-example-5,triaxial-example-6,triaxial-example-7}).  
On the one hand, the majority of the nuclear ground 
states are reflection symmetric. On the other hand, in certain 
regions of the Segr\'e chart, octupole-deformed ground states are present \cite{q3-review}. 
Octupole shapes have already been studied using the macroscopic-microscopic 
(Mac-Mic) model \cite{Mac-Mic-octupole-1,Mac-Mic-octupole-2}, the (mapped) Interacting Boson Model (IBM) 
\cite{mapped-IBM-oct-1,mapped-IBM-oct-2,mapped-IBM-oct-3,mapped-IBM-oct-4,mapped-IBM-oct-5,
mapped-IBM-oct-6,mapped-IBM-oct-7}
as well as microscopic approaches 
\cite{micro-oct-studies-1,micro-oct-studies-2,micro-oct-studies-3,micro-oct-studies-4,
micro-oct-studies-5,micro-oct-studies-6,micro-oct-studies-7,micro-oct-studies-8,
micro-oct-studies-9,micro-oct-studies-10,micro-oct-studies-11,cluster}. Here, one should also
keep in mind the key role played by octupole deformation in the fission paths of
heavy and superheavy nuclei 
\cite{triaxial-example-6,outer-barriers-1,outer-barriers-2}.


The (constrained) Hartree-Fock-Bogoliubov (HFB) approximation \cite{rs} 
is commonly employed, as a  first step, to examine the emergence of 
different types of static deformations in atomic nuclei, by analyzing 
the corresponding mean-field potential energy surfaces (MFPESs). As a 
second step, beyond-mean-field correlations can be accounted for using 
configuration mixing calculations in the spirit of the Generator 
Coordinate Method (GCM) \cite{rs}. Such beyond-mean-field correlations 
determine both the dynamical survival of static  mean-field deformation 
effects and the coupling between different deformation degrees of 
freedom. For example, the dynamical coupling between the  axial 
quadrupole $Q_{20}$ and octupole $Q_{30}$ moments has   been studied  
using the constrained Gogny \cite{gogny} HFB approach, followed by 
two-dimensional (2D) GCM calculations 
\cite{2DGCM-q2q3-Gogny-1,2DGCM-q2q3-Gogny-2, 
2DGCM-q2q3-Gogny-3,2DGCM-q2q3-Gogny-4,2DGCM-q2q3-Gogny-5}. 
Besides the overall agreement with the available experimental data for 
the excitation energies of the lowest negative-parity states as well as 
for the reduced $B(E1)$ and $B(E3)$ strengths,  the Gogny HFB+2D-GCM 
studies \cite{2DGCM-q2q3-Gogny-1,2DGCM-q2q3-Gogny-2, 
2DGCM-q2q3-Gogny-3,2DGCM-q2q3-Gogny-4,2DGCM-q2q3-Gogny-5} also revealed 
a weak $(Q_{20},Q_{30})$-coupling. As a result, in most of the cases, 
key features of the octupole dynamics can be accounted for in 
one-dimensional (1D) GCM calculations, using the octupole moment 
$Q_{30}$ as single collective coordinate. Note that, from a dynamical 
perspective, such weak $(Q_{20},Q_{30})$-coupling translates into the 
alignment of the major axes of the 2D-GCM  collective wave functions 
along either the $Q_{20}$ or $Q_{30}$-axis (see, for example, 
\cite{2DGCM-q2q3-Gogny-5}). The observed weak 
$(Q_{20},Q_{30})$-coupling might be anticipated from the different 
parity quantum numbers of the quadrupole $Q_{20}$ and octupole $Q_{30}$ 
moments. 

The question then naturally arises as to which is the degree 
of coupling between the quadrupole  and the next positive-parity 
multipole moment, i.e., the  hexadecapole $Q_{40}$ moment. Hexadecapole 
shapes, have already received  both theoretical \cite{hexa-prvious-1, 
hexa-prvious-2,hexa-prvious-3,hexa-prvious-4,NL-bb-paper,Nomura-Lotina-b4,Lotina-Sm-Gd-Gogny-mapped,A.Zdeb-scission} 
and experimental \cite{exp-b4-1,exp-b4-2,exp-b4-3,scattering-b2b4} 
attention in the literature. For a previous large scale Gogny-HFB 
survey of static hexadecapole $\beta_{40}$ deformation parameters in 
even-even nuclei, the reader is referred to Ref.~\cite{large-scale-b4}. 
The dynamical 2D-GCM $(\beta_{20}, \beta_{40})$-coupling has also been 
considered in Ref.~\cite{large-scale-b4} for a selected set of Sm and 
Gd isotopes. At variance with the quadrupole-octupole case it has been 
found that, at least for some of the studied isotopes, the principal 
axes of the collective wave functions  align along directions tilted 
with respect to both the $\beta_{20}$ and $\beta_{40}$ axes, i.e., the 
quadrupole and hexadecapole degrees of freedom are  interwoven in those 
nuclei. 

The nontrivial beyond-mean-field physics brought by the 
hexadecapole dynamics, has been further 
considered in a recent Gogny HFB+2D-GCM study of the 
$(\beta_{20}, \beta_{40})$-coupling 
for a selected set of Ra, Th, U and Pu nuclei \cite{Rayner-Robledo-b2-b4-RaThUPu}.
Sizable hexadecapole deformation has been obtained 
for ground and excited states of nuclei around  $^{238}$U, which agrees
well with the  conclusions extracted from the recent analysis
of hydrodynamic simulations of the quark-gluon plasma 
at the BNL Relativistic Heavy Ion Collider (RHIC) \cite{b4-RIHC}. It has  been 
found that a region with small negative hexadecapole deformation, just
below the neutron magic
number $N =184$ \cite{triaxial-example-6}, remains stable once zero-point 
quadrupole-hexadecapole fluctuations are taken 
into account within the 2D-GCM framework. Furthermore, with increasing mass number, a transition 
is predicted \cite{Rayner-Robledo-b2-b4-RaThUPu} from 
a regime in which for both ground and excited states 
the quadrupole and hexadecapole degrees of freedom are interwoven to 
a regime in which they are decoupled. In addition, the previous studies  
\cite{large-scale-b4,Rayner-Robledo-b2-b4-RaThUPu} have 
shown that $(\beta_{20}, \beta_{40})$-GCM
calculations provide a ground state correlation energy gain twice as large 
as the one obtained by considering the quadrupole degree of freedom alone.
This was an important outcome of the study as it proves that the consideration
of the hexadecapole degree of freedom can be necessary to improve our present
description of binding energies all over the nuclear chart.

The previous results indicate that a better understanding of the 
nontrivial beyond-mean-field $(\beta_{20}, \beta_{40})$-coupling in 
other regions of the nuclear chart is still required. With this in 
mind,  we have applied in this work the  HFB+2D-GCM scheme 
\cite{large-scale-b4,Rayner-Robledo-b2-b4-RaThUPu} to study the 
$(\beta_{20}, \beta_{40})$-coupling  in the four isotopic chains 
$^{170-202}$Yb, $^{170-202}$Hf, $^{170-202}$W and $^{170-202}$Os, 
covering sixty eight nuclei. The considered mass range allows  to 
examine not only diamond-like ($\beta_{40} > 0$) but also square-like 
($\beta_{40} < 0$) shapes. It is precisely around the proton  $Z=72$ 
and neutron $N=112$ numbers where, the largest negative static 
hexadecapole deformations have been predicted \cite{large-scale-b4}, a 
few units below the neutron magic number $N=126$. One of our goals in 
this work is precisely to examine those hexadecapole deformation 
effects and their coupling to the quadrupole degree of freedom in this 
region of the nuclear chart. Note that, exception made of the previous 
limited study \cite{Rayner-Robledo-b2-b4-RaThUPu}, a detailed dynamical 
account of negative hexadecapole deformations has not yet been 
presented in the literature. 

At the 2D-GCM level, the present study will examine to which extent the 
quadrupole and hexadecapole degrees of freedom are interwoven in the 
ground and excited states of the considered Yb, Hf, W and Os nuclei. 
Moreover, as will be discussed later on in the paper, the dynamical 
2D-GCM calculations are also motivated by the softness of the MFPESs 
along the quadrupole and/or hexadecapole directions in some of the 
studied isotopes as well as by the enhanced quadrupole shape coexistence 
encountered around the neutron number $N=116$.

The paper is organized as follows. The HFB+2D-GCM framework 
\cite{2DGCM-q2q3-Gogny-1,2DGCM-q2q3-Gogny-2, 
2DGCM-q2q3-Gogny-3,2DGCM-q2q3-Gogny-4,2DGCM-q2q3-Gogny-5,large-scale-b4, 
Rayner-Robledo-b2-b4-RaThUPu} is briefly outlined in Sec.~\ref{Theory}. 
The results obtained within the constrained HFB approach for  
$^{170-202}$Yb, $^{170-202}$Hf, $^{170-202}$W and $^{170-202}$Os are 
discussed in Sec.~\ref{MF_RESULTS}, while 2D-GCM results for those 
nuclei will be presented in Sec.~\ref{GCM_RESULTS}. In particular, 
attention is paid in Sec.~\ref{GCM_RESULTS} to 2D-GCM collective wave 
functions, dynamical deformation effects, correlation energies as well 
as to the comparison with 1D-GCM calculations. Finally, 
Sec.~\ref{conclusions} is devoted to the concluding remarks. 

Let us stress that, both at the HFB and 2D-GCM levels, we have chosen
the parametrization D1S of the Gogny \cite{gogny} energy density 
functional (EDF). Similar qualitative and very often quantitative 
results have been obtained in the same set of nuclei with the 
parametrizations D1M \cite{gogny-d1m} and  D1M$^{*}$ 
\cite{gogny-d1mstar} and, therefore, they will not be explicitly 
discussed in the paper.


\section{Theoretical framework}
\label{Theory}

We first perform  HFB calculations 
with constraints on the axially symmetric 
quadrupole $\hat{Q}_{20}$ and hexadecapole $\hat{Q}_{40}$ operators using the Gogny-D1S \cite{gogny} interaction.  
The quadrupole $Q_{2 0}$ and hexadecapole $Q_{4 0}$ moments 
of the intrinsic HFB state $| \varphi \rangle$
are written as
\begin{equation} \label{def-betas}
Q_{\lambda 0}=
\langle \varphi | \hat{Q}_{\lambda 0} | \varphi \rangle = \frac{3 R_{0}^{\lambda} A}{\sqrt{4 \pi (2\lambda +1)}} \beta_{\lambda 0}
\end{equation}
in terms of the deformation parameters $\beta_{\lambda 0}$  
\cite{large-scale-b4}, with 
$R_{0}=1.2A^{1/3}$ and $A$ the mass number. 
With our convention for the spherical harmonics, the multipole operators in cartesian coordinates are
given by $Q_{20}=z^{2}-\frac{1}{2}r_{\perp}^{2}$ and $Q_{40}=z^{4}-3r_{\perp}^{2}z^{2}-\frac{3}{8}r_{\perp}^{4}$ with 
$r_{\perp}^{2}=x^{2}+y^{2}$. In what follows, for the sake 
of brevity, we will refer to those (axial) parameters $\beta_{\lambda 0}$ simply 
as $\beta_{2}$ and $\beta_{4}$. We will also use the shorthand notation 
$\vec{\beta}= (\beta_{2},\beta_{4})$ \cite{Rayner-Robledo-b2-b4-RaThUPu} to simplify 
some expressions.

To solve the HFB equation an efficient second order gradient method 
\cite{SOGM} is employed and a large axially symmetric 
harmonic oscillator (HO) basis  consisting of 17 major shells is used. The same 
oscillator lengths $b_{\perp}=b_{z}= b_{0}=1.01 A^{1/6}$ are used for 
all the HFB configurations to simplify the computation of overlap 
functions at the GCM level \cite{large-scale-b4, 
Rayner-Robledo-b2-b4-RaThUPu,EWT-1,EWT-2}.

The constrained Gogny-HFB calculations are carried out in a 
$(\beta_{2},\beta_{4})$-grid  with $-0.8 \le \beta_{2} \le 1.2$, $-0.8 
\le \beta_{4} \le 1.2$ and the steps $\delta \beta_{2}= \delta 
\beta_{4}= 0.02$. The reason for such a large set of HFB configurations is 
to cover all possible types of minima (prolate, oblate, etc) 
as well as to pin down regions of abrupt level crossing that might lead to 
sudden changes in the PES. Such large number of HFB 
calculations is made possible due to the robustness and performance of 
our HFB axial solver. Within the large set of HFB configurations we 
choose for the GCM calculation only those with not too large excitation 
energies from the ground state. The GCM calculation is performed with these reduced intervals 
and with a step size of 0.04 in both the $\beta_{2}$ and $\beta_{4}$ cases. Typically,
we are left with 30 configurations in each collective variable, for a total of 900
configurations in the two-dimensional case. In the set of 900 configurations 
there is some amount of linear dependency that is filtered by selecting
only those states in the natural basis (the one diagonalizing the norm matrix) 
with norm eigenvalues larger than $10^{-4}$ \cite{rs}. This procedure leaves us
with around two or three hundred basis states for the two dimensional calculation.

\begin{figure*}
\includegraphics[width=1.0\textwidth]{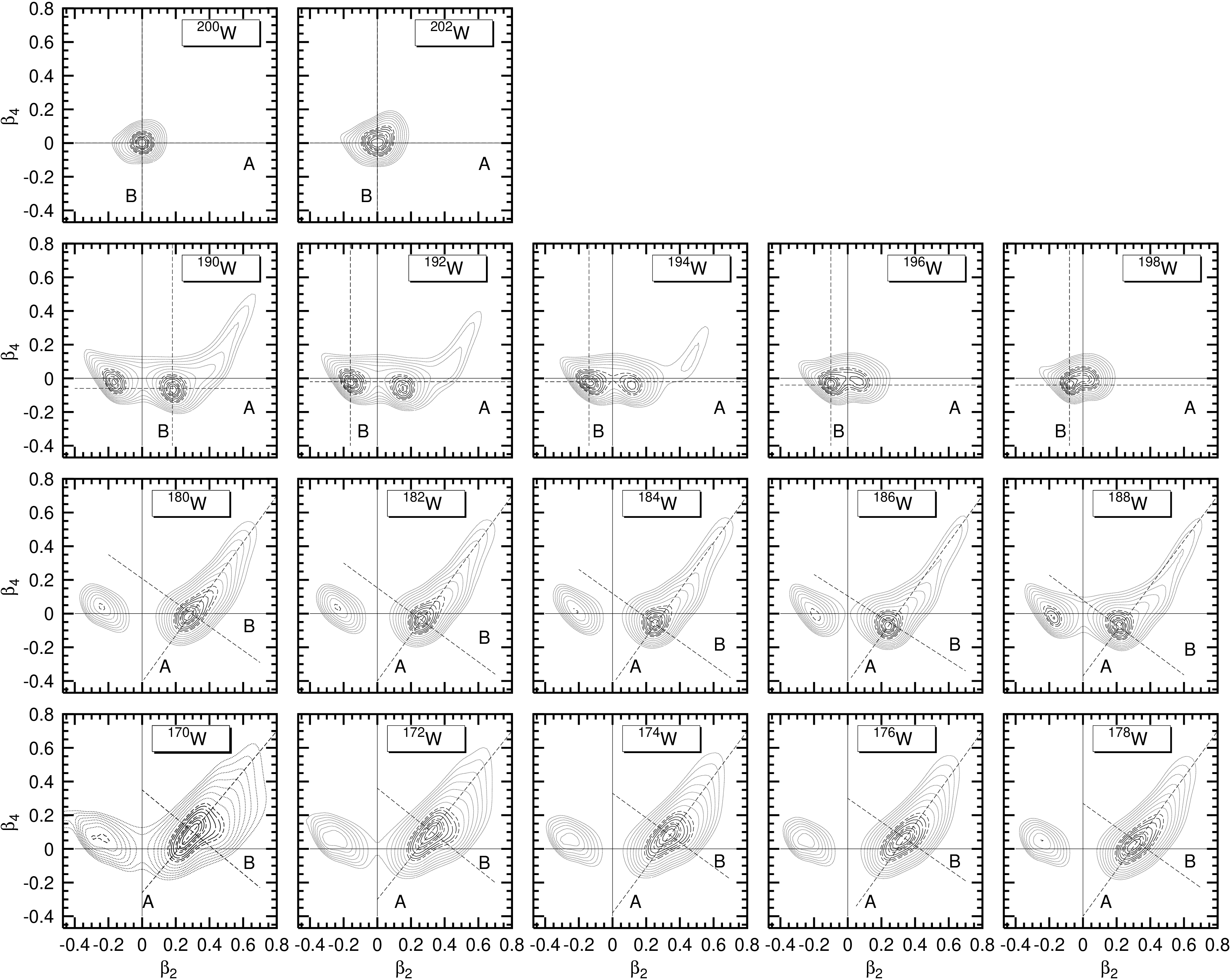}
\caption{MFPESs computed with the Gogny-D1S EDF for the 
isotopes $^{170-202}$W. Contour lines extend from 0.25 MeV up 
to 1 MeV above the ground state energy in steps of 0.25 MeV in the 
ascending sequence full, long-dashed, medium-dashed and short-dashed. 
The next contours following the same sequence correspond to
energies from 1.5 MeV up to 3 MeV above the 
ground state in steps of 0.5 MeV. From there on, 
dotted contour lines are drawn in steps of 1 MeV. For each nucleus, the two 
perpendicular dotted lines A and B are drawn along the principal axes 
of the parabola that approximates the HFB energy around the absolute 
minimum of the MFPES. A vertical full line is drawn to signal the $\beta_{2}=0$
line whereas a full horizontal line is drawn to signal the $\beta_{4}=0$ line.  
For more details, see the main text.
}
\label{mean-field-surfaces-W-1} 
\end{figure*}

As a second step, 2D-GCM calculations based on the ansatz 
\cite{Rayner-Robledo-b2-b4-RaThUPu}
\begin{equation} \label{GCM-WF}
| {\Psi}_{2D-GCM}^{\sigma} \rangle = \int d\vec{\beta} f^{\sigma} (\vec{\beta}) | {\varphi} (\vec{\beta}) \rangle
\end{equation}
are carried out, using the basis states $| {\varphi} (\vec{\beta}) \rangle$ obtained 
in the constrained HFB calculations. In Eq.(\ref{GCM-WF}), the index $\sigma$ numbers
the different GCM solutions. The amplitudes $f^{\sigma} (\vec{\beta})$ are solutions 
of the Griffin-Hill-Wheeler (GHW) equation \cite{rs}
\begin{equation}
{\cal{H}} f^{\sigma} = E^{\sigma} {\cal{N}} f^{\sigma}
\end{equation}
where, ${\cal{N}}(\vec{\beta}_{1},\vec{\beta}_{2})= \langle \varphi(\vec{\beta}_{1})| \varphi(\vec{\beta}_{2}) \rangle$ and
${\cal{H}}(\vec{\beta}_{1},\vec{\beta}_{2})= \langle \varphi(\vec{\beta}_{1})| \hat{H}| \varphi(\vec{\beta}_{2}) \rangle$ represent the norm and Hamiltonian overlaps, respectively.
As in previous studies, in the computation of ${\cal{H}}(\vec{\beta}_{1},\vec{\beta}_{2})$ we 
consider the mixed-density prescription \cite{projden} and use perturbative 
first-order corrections in both the mean value of the proton and neutron 
numbers \cite{2DGCM-q2q3-Gogny-1,2DGCM-q2q3-Gogny-2,
2DGCM-q2q3-Gogny-3,2DGCM-q2q3-Gogny-4,2DGCM-q2q3-Gogny-5,large-scale-b4,Rayner-Robledo-b2-b4-RaThUPu}.

The collective wave functions with a quantum probabilistic interpretation  
\begin{equation} \label{cll-wfs-HW} 
G^{\sigma} (\vec{\beta}_{1}) =   \int d\vec{\beta}_{2} ~
{\cal{N}}^{\frac{1}{2}}(\vec{\beta}_{1},\vec{\beta}_{2}) 
f^{\sigma} (\vec{\beta}_{2})
\end{equation}
are obtained from the HW amplitudes by using the operational square root 
of the norm kernel \cite{rs,Rayner-Robledo-b2-b4-RaThUPu}. 

We are aware that a more general GCM ansatz than the one in
Eq.(\ref{GCM-WF}) might be required to account for the coupling 
to other degrees of freedom in  this region of the nuclear chart. 
In particular, previous studies have found that triaxiality 
and/or $\gamma$-softness can play a role around $^{190}$W
(see, for example, \cite{gamma-190W-1,gamma-190W-2}). 
Certainly, the consideration of triaxial hexadecapole deformations 
will open up the possibility to study $K=4$ bands.
However, such 
a computationally
cumbersome multidimensional 
GCM approach is out of the scope of the present paper in which, we will restrict to the use of the 
axial quadrupole $\beta_{2}$ and 
hexadecapole $\beta_{4}$ deformation parameters as generating coordinates.

Concerning the effect of symmetry restoration on the present results as a 
consequence of including angular momentum and particle number projection,
we do not expect significant qualitative changes as most of the nuclei
considered are well deformed ones with rather strong pairing correlations.
On the other hand, the computational cost of including those symmetry 
restorations will be prohibitively large.

\begin{figure}
\includegraphics[width=0.49\textwidth]{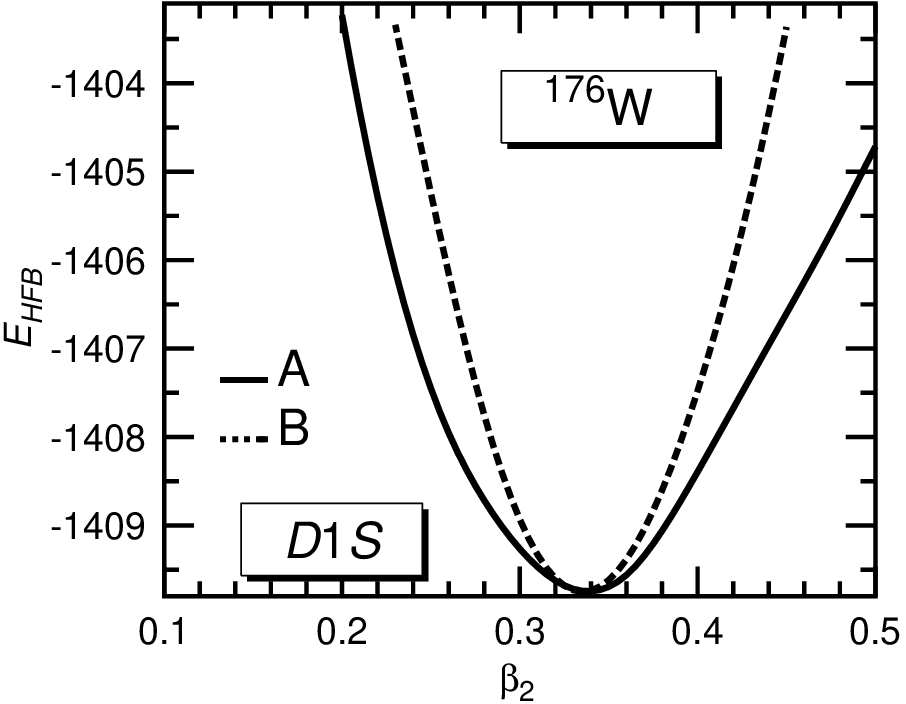}
\caption{HFB energy, as a function of the deformation parameter 
$\beta_{2}$, along  the directions A ($\beta_{4,A}=1.39 \beta_{2} - 0.41$) 
and B ($\beta_{4,B}= -0.71 \beta_{2} + 0.30$) for the nucleus $^{176}$W.
Results are obtained with the Gogny-D1S EDF. For more details, see 
the main text.
}
\label{pathsAB-Energy-176W} 
\end{figure} 


\section{Discussion of the results}
\label{results}

This section is divided in two related contents:
static 
deformation effects are discussed in Sec.~\ref{MF_RESULTS}, while the 
role of beyond-mean-field correlations is considered in 
Sec.~\ref{GCM_RESULTS}.

\subsection{Static deformation effects: HFB calculations}
\label{MF_RESULTS}

The MFPESs obtained for the nuclei $^{170-202}$W are depicted in 
Fig.\ref{mean-field-surfaces-W-1}, as illustrative examples. Similar 
results are obtained for $^{170-202}$Yb, $^{170-202}$Hf and 
$^{170-202}$Os. As can be seen from the figure, for the isotopes
$^{170-190}$W the absolute minima of the  MFPESs 
correspond to $\beta_{2} > 0$ values within the range
$0.18 \le \beta_{2} \le 0.34$. The largest (smallest) prolate
deformation $\beta_{2}= 0.34$ ($0.18$) corresponds
to $^{174,176}$W ($^{190}$W). On the other hand, oblate-deformed
ground states are predicted for $^{192-198}$W with 
$\beta_{2}= -0.16, -0.14, -0.10$ and $-0.08$. A 
minimum with $\beta_{2}= 0$ is obtained for $^{200,202}$W.

Shape transitions are also observed along the $\beta_{4}$-direction. 
Diamond-like shapes are predicted for $^{170-178}$W with the associated  
$\beta_{4}$ = 0.10, 0.10, 0.08, 0.06 and 0.04 values.  Consistent with 
the expectations within the polar-gap model 
\cite{polar-gap-1,polar-gap-2}, for all the studied isotopic chains a 
region with negative hexadecapole deformations, i.e., square-like 
shapes, is obtained just below the neutron magic number N=126. For 
example, for $^{182-198}$W we obtain $\beta_{4}$ = -0.04, -0.06, 
-0.08, -0.08, -0.06, -0.02, -0.02, -0.04 and -0.04, respectively. On 
the other hand, the HFB ground states  of  $^{180,200,202}$W correspond 
to $\beta_{4} = 0$.

Figure~\ref{mean-field-surfaces-W-1} also provides a (static) hint on 
the coupling between the quadrupole and hexadecapole degrees in the 
considered nuclei. For each nucleus, the dotted line A runs parallel to 
the bottom of the energy valley in the neighborhood of the absolute 
minimum of the MFPES. On the other hand, the dotted line B represents 
the direction perpendicular to A. For $^{170-188}$W, the lines A and B 
are tilted with respect to both the $\beta_{2}$ and $\beta_{4}$ axes. 
Along both directions A and B, the parameter $\beta_{4}$ exhibits a 
linear dependence on $\beta_{2}$. For example, for $^{176}$W  we 
obtain the parametrization $\beta_{4,A}=1.39 \beta_{2} - 0.41$ and 
$\beta_{4,B}= -0.71 \beta_{2} + 0.30$, using configurations around the 
absolute minimum of the corresponding MFPES.

From Fig.\ref{pathsAB-Energy-176W} one realizes that, in the 
neighborhood of the absolute minimum of the MFPES, the energy displays 
a parabolic behavior along the directions A and B. As already noted in 
previous works \cite{large-scale-b4,Rayner-Robledo-b2-b4-RaThUPu}, 
these features imply that 1D-GCM calculations, using the deformation 
parameters $\beta_{2}$ or $\beta_{4}$, as single generating coordinates 
essentially scan the same configurations around the minimum of the 
MFPESs in $^{170-188}$W, i.e., the quadrupole and hexadecapole degrees 
are interwoven in those isotopes and, therefore, full-fledged 2D-GCM 
calculations must be carried out. For the heavier isotopes 
$^{190-202}$W the directions A and B run parallel to the $\beta_{2}$ 
and $\beta_{4}$ axes, respectively, pointing towards a decoupling of 
the quadrupole and hexadecapole  degrees of freedom.  

\begin{figure*}
\includegraphics[width=1.0\textwidth]{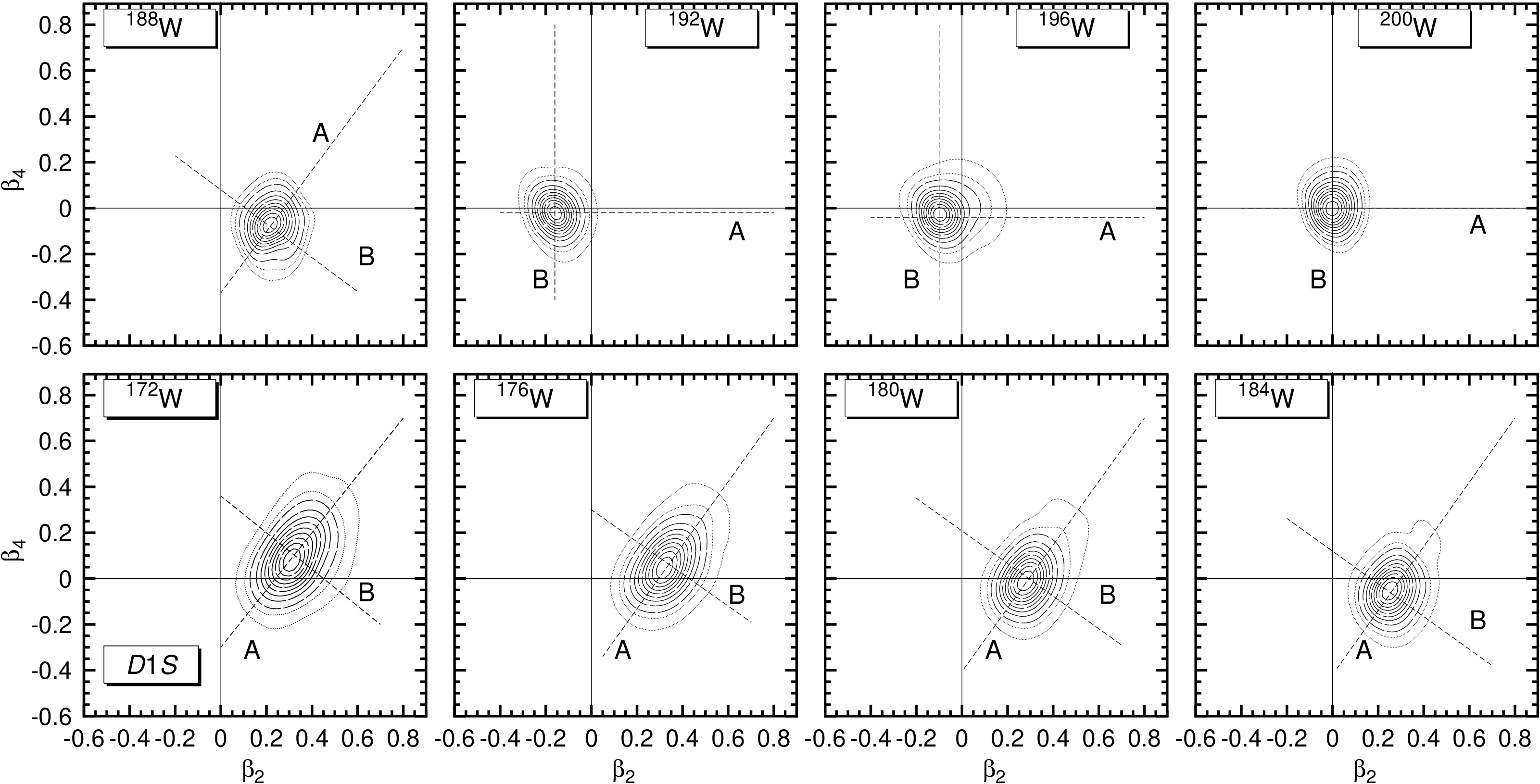}
\caption{Collective wave functions Eq.(\ref{cll-wfs-HW}) corresponding 
to the ground states of W isotopes with A=172, 176, 180, 184, 188, 192, 196, 
and 200. The succession of solid, long dashed and short dashed contour 
lines starts at 90$\%$ of the maximum value up to 10$\%$ of it. The two 
dotted-line contours correspond to the tail of the amplitude (5$\%$ and 
1$\%$ of the maximum value).  For each nucleus, the two perpendicular 
dotted lines A and B are drawn along the principal axes of the parabola 
that approximates the HFB energy around the absolute minimum of the 
MFPES and are the same as in Fig.\ref{mean-field-surfaces-W-1}. A 
vertical full line is drawn to signal the $\beta_{2}=0$ line whereas a 
full horizontal line is drawn to signal the $\beta_{4}=0$ line. Results 
are obtained with the Gogny-D1S EDF. For more details, see the 
main text.
}
\label{GSCWF-W} 
\end{figure*}

From the previous Gogny-HFB results we conclude that, for the studied 
Yb, Hf, W and Os chains, a transition occurs with increasing mass 
number from a regime in which the quadrupole and hexadecapole degrees 
of freedom are interwoven to a regime in which they are decoupled. In 
particular, at the static HFB level, the nuclei $^{170-186}$Yb, 
$^{170-188}$Hf, $^{170-188}$W and $^{170-188}$Os belong to the 
($\beta_{2},\beta_{4}$)-coupled regime. This feature by itself calls 
for a full-fledged  2D-GCM analysis including beyond-mean-field 
$(\beta_{2},\beta_{4})$-fluctuations. In addition to the quadrupole and 
hexadecapole HFB shape transitions already discussed above, an enhanced 
shape coexistence is predicted in the HFB calculations around the N=116 
isotope $^{190}$W. For example, in the case of $^{186,188,190,192}$W, 
the configurations located at the $(\beta_{2},\beta_{4})$-coordinates 
$(-0.20,0.00)$, $(-0.18,0.00)$, $(-0.16,-0.02)$ and $(0.16,-0.06)$ lie 
2.65, 1.42, 0.32 and 0.44 MeV, respectively, above the corresponding 
ground states. Furthermore, for some of the studied nuclei, the MFPESs 
display a soft behavior along the quadrupole and/or  hexadecapole 
directions. All these features indicate that configuration mixing 
$(\beta_{2},\beta_{4})$-GCM calculations are required for the 
considered nuclei. The results of those 2D-GCM calculations will be 
discussed in the next Sec.\ref{GCM_RESULTS}.

\subsection{Beyond-mean-field correlations: 2D-GCM calculations}
\label{GCM_RESULTS}

The 2D-GCM collective wave functions corresponding to the ground states 
for a selected set of W isotopes are depicted in Fig.~\ref{GSCWF-W}, to 
illustrate the main features of the results. It should be noted that 
the peaks of the collective wave functions $G^{\sigma=1} (\vec{\beta})$  
correspond to  ($\beta_{2},\beta_{4}$)-deformations rather close to the 
ones obtained at the minimum of the HFB energy (see, 
Fig.\ref{mean-field-surfaces-W-1}). For $^{172,176,180,184,188}$W the 
amplitudes $G^{\sigma=1} (\vec{\beta})$ align along the direction A, 
which is tilted with respect to both the quadrupole and hexadecapole 
axes. For the heavier isotopes  $^{192,196,200}$W,  the amplitudes 
$G^{\sigma=1} (\vec{\beta})$ align parallel to the quadrupole axis. 
From our analysis of the collective wave functions $G^{\sigma=1} 
(\vec{\beta})$ obtained for the Yb, Hf, W and Os nuclei, we conclude 
that the quadrupole and hexadecapole degrees of freedom are interwoven 
in the ground states up to the mass number $A=186-188$.

\begin{figure*}
\includegraphics[width=1.00\textwidth]{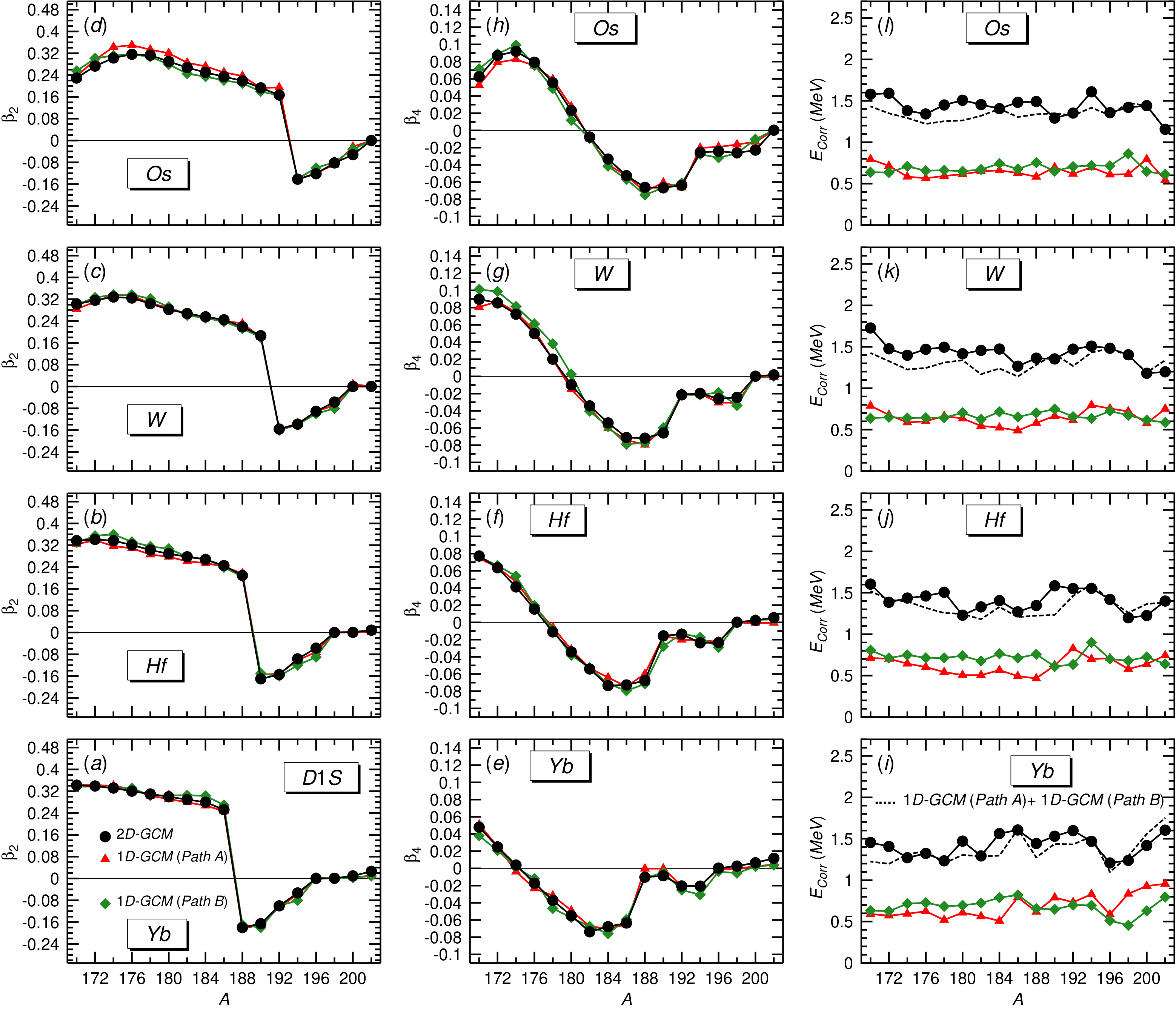}
\caption{(Color online) 2D-GCM ground state quadrupole $\beta_{2}$ 
[panels (a)-(d)] and hexadecapole $\beta_{4}$ [panels (e)-(h)] 
deformation parameters for $^{170-202}$Yb, $^{170-202}$Hf, 
$^{170-202}$W and $^{170-202}$Os. The 2D-GCM correlation energies 
$E_{Corr}$ obtained for those nuclei are depicted in panels (i)-(l). 
Quadrupole and hexadecapole deformations as well as correlation 
energies obtained in 1D-GCM calculations along the directions A and B 
are also included in the plots. The dotted lines in panels (i)-(l) 
correspond to the sum of 1D-GCM correlation energies along the 
directions A and B. Results are obtained with the Gogny-D1S EDF. 
For more details, see the main text. 
}
\label{2DGCM-b2-b4} 
\end{figure*} 

Having the collective wave functions $G^{\sigma} (\vec{\beta})$ at 
hand, we compute \cite{Rayner-Robledo-b2-b4-RaThUPu} the 
dynamical deformation parameters $\beta_{\lambda,2D-GCM}^{\sigma}$, 
with $\lambda = 2,4$. The ground state quadrupole 
$\beta_{2,2D-GCM}^{\sigma=1}$  parameters obtained for $^{170-202}$Yb, 
$^{170-202}$Hf, $^{170-202}$W and $^{170-202}$Os are plotted in panels 
(a)-(d) of Fig.~\ref{2DGCM-b2-b4}. The nuclei $^{170-186}$Yb, 
$^{170-188}$Hf, $^{170-190}$W and $^{170-192}$Os exhibit prolate 2D-GCM 
ground states, with the largest $\beta_{2,2D-GCM}^{\sigma=1}$ values  
in the range [0.32,0.34] corresponding  to $^{170,172}$Yb, 
$^{170-74}$Hf, $^{174}$W and $^{176,178}$Os. With increasing mass 
number, the $\beta_{2,2D-GCM}^{\sigma=1}$ deformations decrease 
smoothly up to $\beta_{2,2D-GCM}^{\sigma=1}=$ 0.25, 0.21, 0.19 and 0.17 
for $^{186}$Yb, $^{188}$Hf, $^{190}$W and $^{192}$Os. Oblate ground 
states are obtained for $^{188-194}$Yb, $^{190-196}$Hf, 
$^{192-198}$W and $^{194-200}$Os. For the N=126 nuclei $^{196}$Yb, 
$^{198}$Hf, $^{200}$W and $^{202}$Os as well as for their heavier 
neighbors we obtain $\beta_{2,2D-GCM}^{\sigma=1} \approx 0$.

The hexadecapole $\beta_{4,2D-GCM}^{\sigma=1}$  parameters are 
displayed in panels (e)-(h) of Fig.\ref{2DGCM-b2-b4}. For the lighter 
Yb, Hf and W isotopes diamond-like shapes with decreasing  
$\beta_{4,2D-GCM}^{\sigma=1}$ values are obtained. This is also the 
case for Os isotopes after reaching the largest positive value  
$\beta_{4,2D-GCM}^{\sigma=1}=0.09$ for $^{176}$Os. The 2D-GCM 
calculations confirm the dynamical survival of a region (i.e., the 
nuclei $^{176-194}$Yb, $^{178-196}$Hf, $^{180-198}$W and 
$^{182-200}$Os) characterized by square-like ground state shapes, with 
$-0.07 \le \beta_{4,2D-GCM}^{\sigma=1} \le -0.01$, just below the 
neutron magic number N=126,  Note that, the largest 
$\beta_{4,2D-GCM}^{\sigma=1} < 0$ values correspond to the nuclei 
$^{182}$Yb, $^{184}$Hf, $^{188}$W  and $^{190}$Os. 

\begin{figure*}
\includegraphics[width=1.0\textwidth]{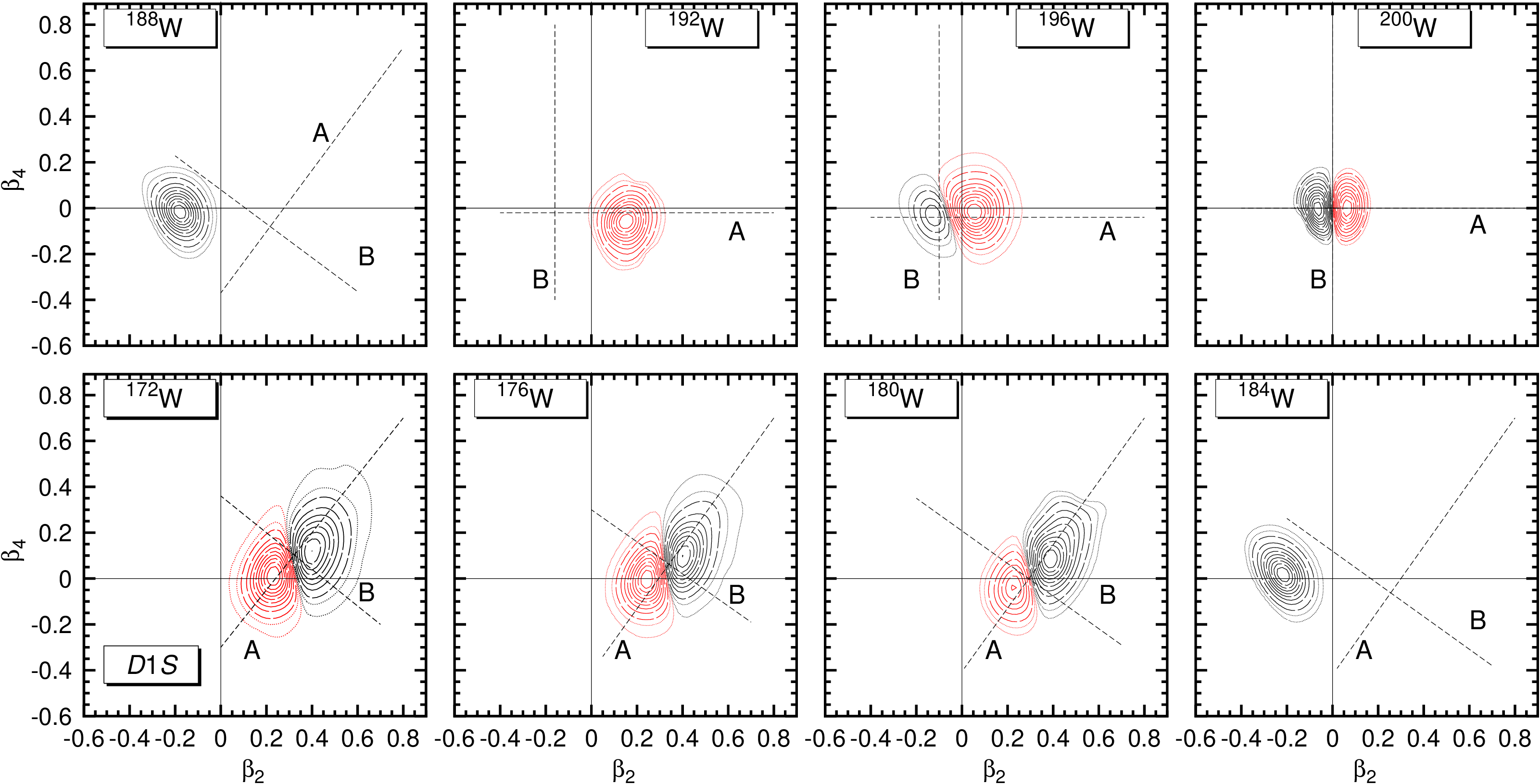}
\caption{(Color online) The same as in Fig.~\ref{GSCWF-W} but for the first excited 
states of the nuclei $^{172,176,180,184,188, 192, 196, 200}$W. Red 
contour lines correspond to negative values of the collective wave functions.
Please note that in $^{184-192}$W there is shape coexistence and the first
excited state corresponds to a "ground state like" wave function but 
centered in the excited minimum.
}
\label{FEXCCWF-W} 
\end{figure*}

The previous results agree well with the ones expected within the polar 
gap model \cite{polar-gap-1,polar-gap-2}. A region with ground state 
square-like shapes, just below the neutron magic number N=184, has also 
been predicted in previous 2D-GCM calculations for Ra, Th, U and Pu 
isotopes \cite{Rayner-Robledo-b2-b4-RaThUPu}. Furthermore, our 2D-GCM 
calculations confirm previous (static) Gogny-HFB results 
\cite{large-scale-b4} indicating that the largest negative hexadecapole 
deformations, all over the nuclear chart, are found around the proton 
Z=72 and neutron N=112 numbers. Let us also stress that, though not 
explicitly discussed in this paper, very similar results are
obtained in 2D-GCM calculations with the parametrization D1M 
\cite{gogny-d1m} and  D1M$^{*}$ \cite{gogny-d1mstar} of the Gogny-EDF. 
This clearly shows that the origin of the discussed effects is 
independent of the interaction used to describe them.

The  correlation energies $E_{Corr, 2D-GCM}$, defined 
as the difference between the
HFB and 2D-GCM ground state energies, are plotted in panels (i)-(l)
of Fig.\ref{2DGCM-b2-b4}. For the studied  isotopic
chains, we obtain  values within the range
$1.15~\textrm{MeV} \le E_{Corr, 2D-GCM} \le 1.73~\textrm{MeV}$. Note, that this range 
of variation  ($0.58~MeV$) is comparable 
with typical  rms deviations  for the binding 
energy in the case of the  parametrizations  D1M 
\cite{gogny-d1m}
or D1M$^{*}$ \cite{gogny-d1mstar} of the Gogny-EDF.
Thus, we conclude that the correlation 
energies $E_{Corr, 2D-GCM}$ should be taken into account 
in future parametrizations of the Gogny-EDF
\cite{large-scale-b4,Rayner-Robledo-b2-b4-RaThUPu}
.

In addition to the 2D-GCM calculations, we carry out 1D-GCM 
calculations using the quadrupole parameter $\beta_{2}$ as single 
generating coordinate. In those $\beta_{2}$-GCM calculations, we 
employ the grid $-0.8 \le \beta_{2} \le 1.2$ with the step $\delta 
\beta_{2}= 0.01$. We obtain quadrupole correlation energies 
within the range $0.49~\textrm{MeV} \le E_{Corr, 1D-GCM, \beta_{2}} \le 
0.97~\textrm{MeV}$. Thus, in going from the $\beta_{2}$-GCM to the 
$(\beta_{2},\beta_{4})$-GCM calculations we obtain an additional 
correlation energy in the range $0.66~\textrm{MeV} \le \delta E_{Corr} 
\le 0.76~\textrm{MeV}$ due to hexadecapole effects which is comparable 
to the values of the quadrupole correlation energy itself. These 
results and the ones obtained in previous studies 
\cite{2DGCM-q2q3-Gogny-1,2DGCM-q2q3-Gogny-2, 
2DGCM-q2q3-Gogny-3,2DGCM-q2q3-Gogny-4,2DGCM-q2q3-Gogny-5,large-scale-b4, 
Rayner-Robledo-b2-b4-RaThUPu} represent a warning for GCM practitioners 
regarding the slow convergence of the nuclear correlation energies with 
respect to the number of (even and/or odd) deformation parameters 
employed in the GCM ansatz and deserve further investigation. Work 
along these lines is in progress and 
 will be reported in future publications.

We also perform 1D-GCM calculations along the directions A and 
B. Those directions are characterized by the linear dependencies 
$\beta_{4,A}=a_{A} \beta_{2} - b_{A}$ and $\beta_{4,B}=-(1/a_{A}) 
\beta_{2} + b_{B}$, respectively. As already discussed in 
Sec.~\ref{MF_RESULTS}, for each of the considered nuclei, the 
parameters $a_{A}$, $b_{A}$ and $b_{B}$ are fitted, using 
configurations around the absolute minimum of the corresponding MFPES. 
In those A and B 1D-GCM calculations, we use the 
$\beta_{2}$-grid $-0.8 \le \beta_{2} \le 1.2$ with the step $\delta 
\beta_{2}= 0.01$. 

As can be seen from panels (a)-(d) and (e)-(h) of 
Fig.\ref{2DGCM-b2-b4}, the quadrupole and hexadecapole deformations 
obtained in the A and B 1D-GCM calculations are rather similar to the 
corresponding $\beta_{2,2D-GCM}^{\sigma=1}$ and 
$\beta_{4,2D-GCM}^{\sigma=1}$ values. The correlation energies 
$E_{Corr,1D-GCM,A}$ and $E_{Corr,1D-GCM,B}$ are also included in panels 
(i)-(l) of the figure. As expected from previous discussions, the 
energies $E_{Corr,1D-GCM,A}$ and $E_{Corr, 1D-GCM, \beta_{2}}$ are 
similar. From panels (i)-(l), one realizes that the sum 
$E_{Corr,1D-GCM,A}+ E_{Corr,1D-GCM,B}$  accounts for a significant 
portion of the total 2D-GCM correlation energy $E_{Corr, 2D-GCM}$. 
 
Let us stress that, in going from  either the A or B 1D-GCM  to the 
2D-GCM calculations, we obtain an additional correlation energy in the 
range  $0.68~\textrm{MeV} \le \delta E_{Corr} \le 0.77~\textrm{MeV}$ 
and $0.69~\textrm{MeV} \le \delta E_{Corr} \le 0.87~\textrm{MeV}$, 
respectively. Those additional  correlation energies are also 
comparable to the quadrupole correlation energy itself. The comparison 
of the $\beta_{2}$-GCM, A-GCM, B-GCM and 2D-GCM correlation energies 
obtained for the studied Yb, Hf, W and Os isotopes, as well as previous 
results for Sm, Gd, Ra, Th, U and Pu nuclei 
\cite{large-scale-b4,Rayner-Robledo-b2-b4-RaThUPu}, reveal the 
nontrivial role of beyond-mean-field zero-point 
$(\beta_{2},\beta_{4})$-fluctuations in the ground states of nuclei in 
different  regions of the nuclear chart.

\begin{figure*}
\includegraphics[width=1.00\textwidth]{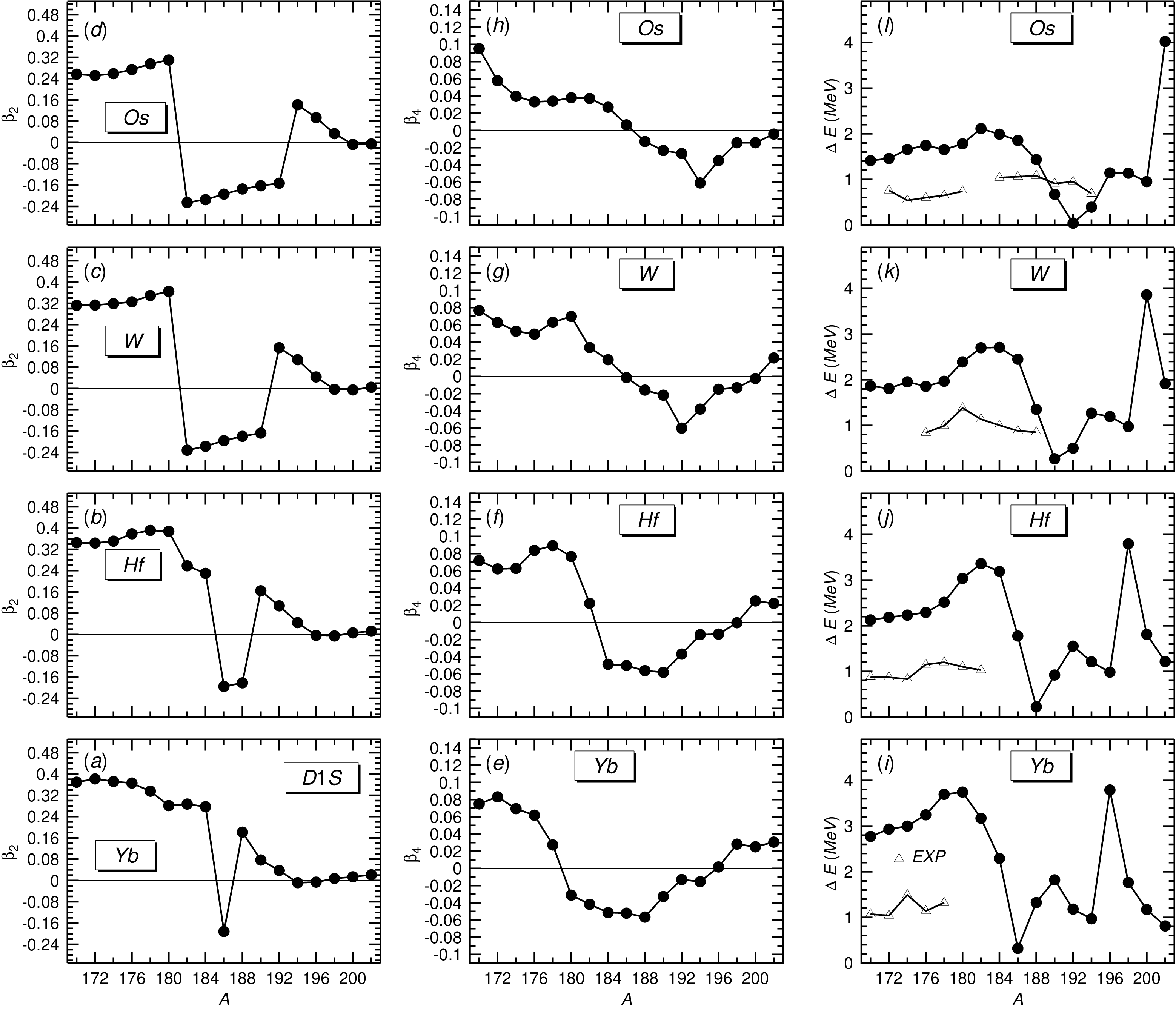}
\caption{2D-GCM quadrupole $\beta_{2}$ [panels (a)-(d)] and 
hexadecapole $\beta_{4}$ [panels (e)-(h)] deformation parameters 
corresponding to the first excited states in $^{170-202}$Yb, 
$^{170-202}$Hf, $^{170-202}$W and $^{170-202}$Os. The excitation 
energies $\Delta E^{ \sigma=2}$ corresponding to those states are 
depicted in panels (i)-(l). Experimental data taken from \cite{exp} 
is also shown as open triangles for comparison. See text for a discussion
of the theory-experiment comparison. Results are obtained with the 
Gogny-D1S EDF. For more details, see the main text. 
}
\label{2DGCM-b2-b4-FEXC} 
\end{figure*}
 
The 2D-GCM collective wave functions corresponding to the first excited 
states in  $^{172,176,180,184,188, 192, 196, 200}$W are depicted in 
Fig.~\ref{FEXCCWF-W}, as illustrative examples. For the isotopes 
$^{172,176,180}$W, the functions $G^{\sigma=2} (\vec{\beta})$ 
correspond to phonons aligned along the tilted direction A. In the case 
of the heavier isotopes $^{196, 200}$W, those collective wave functions 
align along the $\beta_{2}$-axis and represent pure two-phonon 
quadrupole vibrations. The 2D-GCM ground state of $^{184,188}$W 
($^{192}$W) exhibits prolate (oblate) deformation. Therefore, given the 
enhanced shape coexistence around the neutron number N=116 (see, 
Sec.~\ref{MF_RESULTS}), it is not surprising that the collective 
strength $G^{\sigma=2} (\vec{\beta})$ in $^{184,188}$W ($^{192}$W) 
corresponds to a shape isomeric state and concentrates on the oblate 
(prolate) side. We thus observe that the  structural evolution of the 
collective wave functions $G^{\sigma=2} (\vec{\beta})$ obtained not 
only for W  but also for Yb, Hf and Os nuclei also points towards a 
transition from a coupled to a decoupled ($\beta_{2},\beta_{4}$)-regime 
with increasing mass number. In particular, for all the studied 
isotopic chains, the quadrupole and hexadecapole degrees of freedom are 
interwoven in the 2D-GCM first excited states up to the mass number 
A=184-188.

The quadrupole  $\beta_{2,2D-GCM}^{\sigma=2}$ parameters obtained for  
the studied Yb, Hf, W and Os nuclei are plotted in panels (a)-(d) of 
Fig.\ref{2DGCM-b2-b4-FEXC}. The first excited states of $^{170-184}$Yb, 
$^{170-184}$Hf, $^{170-180}$W and $^{170-180}$Os exhibit prolate 
deformations within the range $0.25 \le \beta_{2,2D-GCM}^{\sigma=2} \le 
0.39$. On the other hand, oblate first excited states are predicted for 
$^{186}$Yb and  $^{186,188}$Hf. For both the W and Os isotopic chains, 
an enlarged region with neutron numbers $106 \le N \le 116$ corresponds 
to $\beta_{2,2D-GCM}^{\sigma=2} < 0$ values. These quadrupole shape 
transitions, encoded in the structure of the corresponding 
$G^{\sigma=2} (\vec{\beta})$ collective strengths, reflect the soft 
behavior of the MFPESs (see, for example, 
Fig.~\ref{mean-field-surfaces-W-1}) as well as the enhanced competition 
between  low-lying configurations, based on different intrinsic 
deformations, as we approach the neutron number N=116. Note that, as 
already illustrated in Fig.~\ref{FEXCCWF-W} for the case of W isotopes,  
the subtle balance between competing configurations leads, in some 
instances, to a concentration of a significant portion of the 
$G^{\sigma=2} (\vec{\beta})$ collective wave functions on the oblate 
side. Prolate deformations $\beta_{2,2D-GCM}^{\sigma=2}= 0.14-0.18$ are 
obtained once more for $^{188}$Yb, $^{190}$Hf, $^{192}$W and 
$^{194}$Os, while for larger mass numbers those deformations decrease 
up to $\beta_{2,2D-GCM}^{\sigma=2} \approx 0$.

The hexadecapole  $\beta_{4,2D-GCM}^{\sigma=2}$ parameters 
corresponding to the first excited states in $^{170-202}$Yb, 
$^{170-202}$Hf, $^{170-202}$W and $^{170-202}$Os are depicted in panels 
(e)-(h) of Fig.\ref{2DGCM-b2-b4-FEXC}. From a dynamical perspective the 
results clearly indicate that hexadecapole deformations play a role not 
only in the ground (see, Fig.\ref{2DGCM-b2-b4}) but also in the first 
excited states of nuclei in this region of the nuclear chart. In 
particular, the first excited states of $^{170-178}$Yb, $^{170-182}$Hf, 
$^{170-184}$W and $^{170-186}$Yb correspond to diamond-like shapes. 
However, for all the studied isotopic chains, a region with square-like 
first excited states is found in the calculations just below the 
neutron magic number N=126. Note, that the largest (negative) 
$\beta_{4,2D-GCM}^{\sigma=2} \approx -0.06$ values correspond to the 
N=118 isotones $^{188}$Yb, $^{190}$Hf, $^{192}$W and $^{194}$Os. We 
stress that this is the region where an enhanced shape coexistence is 
predicted in the calculations.

The excitation energies $\Delta E^{\sigma = 2}$ are shown in panels 
(i)-(l) of Fig.\ref{2DGCM-b2-b4-FEXC}. They reach their lowest values 
$\Delta E^{\sigma = 2}= 320, 223, 268$ and $43$ keV for  $^{186}$Yb, 
$^{188}$Hf, $^{190}$W and $^{192}$Os, respectively. It is precisely the 
enhanced shape coexistence what leads to such low  values of the 
energies $\Delta E^{\sigma = 2}$ around the neutron number N=116. On 
the other hand, the pronounced maxima observed in the energies $\Delta 
E^{\sigma = 2}$ obtained for  $^{196}$Yb, $^{198}$Hf, $^{200}$W and 
$^{202}$Os represent the strong signature of the N=126 neutron shell 
closure. We have included experimental data about the first $0^{+}$
state in those nuclei in the figure. However, the nature of the 
measured $0^{+}$ is still under vivid debate and it is not clear if
a comparison with our results that include shape coexistence, $\beta_{2}$
and $\beta_{4}$ $K=0$ vibrations for the first excited state is meaningful.

\begin{figure*}
\includegraphics[width=1.00\textwidth]{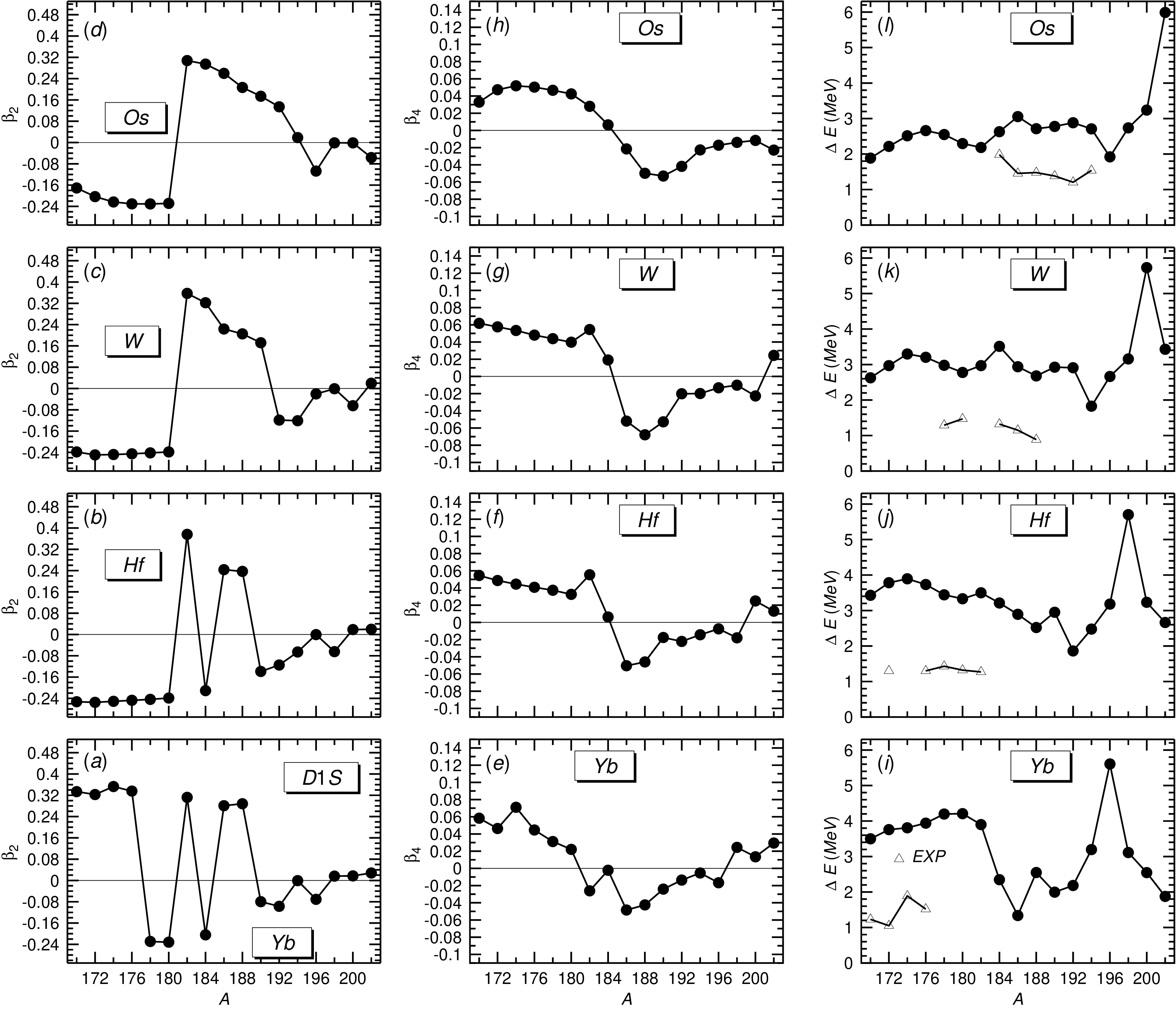}
\caption{2D-GCM quadrupole $\beta_{2}$ [panels (a)-(d)] and 
hexadecapole $\beta_{4}$ [panels (e)-(h)] deformation parameters 
corresponding to the second excited states in $^{170-202}$Yb, 
$^{170-202}$Hf, $^{170-202}$W and $^{170-202}$Os. The excitation 
energies $\Delta E^{ \sigma=2}$ corresponding to those states are 
depicted in panels (i)-(l). Experimental data taken from \cite{exp} 
is also shown as open triangles for comparison. Results are obtained with the 
Gogny-D1S EDF. For more details, see the main text. 
}
\label{2DGCM-b2-b4-2nd} 
\end{figure*}

Finally, let us briefly comment on the results we
obtain for the
second excited states in the Yb, Hf, W and Os chains. In this case, the 
structural evolution of the collective wave functions
$G^{\sigma=3} (\vec{\beta})$ once more reveals that the quadrupole and hexadecapole degrees of freedom 
are coupled up to the mass number A=184-188. For example, for the same 
W isotopes illustrated in Figs.~\ref{GSCWF-W} and \ref{FEXCCWF-W}, we obtain
that the collective wave functions
$G^{\sigma=3} (\vec{\beta})$ evolve from oblate 
($\beta_{2,2D-GCM}^{\sigma=3} \approx  -0.24$)
configurations
in $^{172,176,180}$W
to a prolate 
($\beta_{2,2D-GCM}^{\sigma=3}= 0.32)$
phonon aligned along the direction A in $^{184}$W. In the case of $^{188}$W, the 
second excited state corresponds to a prolate
($\beta_{2,2D-GCM}^{\sigma=3}= 0.21)$
phonon aligned along the direction B, while 
for $^{192,196,200}$W  we obtain oblate 
($\beta_{2,2D-GCM}^{\sigma=3}= -0.12, -0.02$ and $-0.06$)
two phonon excitations 
aligned along the $\beta_{2}$-axes. Second excited diamond-like 
(square-like) states
with $0.02 \le \beta_{4,2D-GCM}^{\sigma=3} \le 0.07$
($-0.06 \le \beta_{4,2D-GCM}^{\sigma=3} \le -0.01$ )
are found 
for $^{170-180}$Yb, $^{170-182}$Hf, $^{170-184}$W and 
$^{170-182}$Os ($^{182-196}$Yb, $^{186-198}$Hf, $^{186-200}$W and 
$^{186-202}$Os). The excitation energies $\Delta E^{\sigma = 3}$ reach their 
lowest values 
$1.34, 1.86, 1.83$ and $1.92$ MeV  for 
$^{186}$Yb, $^{192}$Hf, $^{194}$W and $^{196}$Os. Those energies also
exhibit a pronounced peak for the N=126 isotones. A summary of the results is shown in 
Fig.~\ref{2DGCM-b2-b4-2nd}.

\section{Conclusions}
\label{conclusions}

In this work, the roles of hexadecapole deformations and 
quadrupole-hexadecapole configuration mixing at the static (HFB) and 
dynamical (2D-GCM) levels for Yb, Hf, W and Os isotopes within the mass 
range $170 \le A \le 202$ are considered. To this end, we consider
the parametrization D1S of the Gogny-EDF both at the mean-field 
level and beyond. 

At the static HFB level, for each of the studied chains, the analysis 
of the  MFPESs shows the emergence of global minima corresponding to 
hexadecapole deformations different from zero for several isotopes. The key role 
played by dynamical hexadecapole deformations in the ground and excited 
states of the considered  nuclei is  confirmed at the 2D-GCM level. 
While the 2D-GCM ground and excited states of the lighter Yb, Hf, W and 
Os isotopes are associated with diamond-like shapes, with increasing 
mass number a region where those states correspond to square-like 
shapes is found in the calculations, just below the neutron shell 
closure N=126. Our 2D-GCM calculations also confirm previous (static) 
Gogny-HFB results \cite{large-scale-b4} indicating that the largest 
negative hexadecapole deformations, all over the nuclear chart, are 
found around the proton Z=72 and neutron N=112 numbers.

The structural evolution of the 2D-GCM collective wave functions 
associated with the ground and excited states  also reveals that, with 
increasing mass number A, a transition occurs from a regime in which 
the quadrupole and hexadecapole degrees of freedom are interwoven to a 
regime in which they are decoupled. For the studied Yb, Hf, W and Os 
nuclei it is found that the ground and excited states belong to the 
coupled regime up to  A=184-188. Such structural evolution is also 
accompanied by an enhanced shape coexistence around the neutron number 
N=116 that leads to the lowest values of  the excitation energies of 
the  second and third 2D-GCM states at N=116 and N=120.

In good agreement with previous studies 
\cite{large-scale-b4,Rayner-Robledo-b2-b4-RaThUPu}, the range of 
variation (0.58~MeV) of the 2D-GCM correlation energies obtained for 
the studied Yb, Hf, W and Os nuclei compares well with the rms 
deviation for the binding energy in recent parametrizations 
\cite{gogny-d1m,gogny-d1mstar} of the Gogny-EDF, and suggests that 
those energies should be included in future fitting protocols of the 
functional. Furthermore, the comparison between 2D-GCM and several 
types of 1D-GCM calculations also corroborates 
\cite{large-scale-b4,Rayner-Robledo-b2-b4-RaThUPu} that the inclusion 
of hexadecapole deformation in the ground state dynamics brings a non 
trivial additional correlation energy comparable to the quadrupole 
correlation energy itself, and encourages a more detailed analysis of 
the slow convergence of the nuclear correlation energies.  Work along 
these lines is in progress and will be reported in future publications.

\begin{acknowledgments}
This research is funded by Nazarbayev University under  Faculty 
Development Competitive Research Grants Program (FDCRGP) for 2025-2027 
Grant 040225FD4712, R. Rodr\'{\i}guez-Guzm\'an. The work of LMR is 
supported by Spanish Agencia Estatal de Investigacion (AEI) of the 
Ministry of Science and Innovation under Grant No. 
PID2021-127890NB-I00. 
\end{acknowledgments}

\end{document}